\documentclass[%
 aip,
 amsmath,amssymb,
 reprint,%
]{revtex4-2}

\usepackage{graphicx}
\usepackage{dcolumn}
\usepackage{bm}
\usepackage{color}

\usepackage[utf8]{inputenc}
\usepackage[T1]{fontenc}
\usepackage{mathptmx}
\usepackage{etoolbox}
\usepackage{mathtools}
\usepackage{scalerel}

\makeatletter
\def\@email#1#2{%
 \endgroup
 \patchcmd{\titleblock@produce}
  {\frontmatter@RRAPformat}
  {\frontmatter@RRAPformat{\produce@RRAP{*#1\href{mailto:#2}{#2}}}\frontmatter@RRAPformat}
  {}{}
}%
\makeatother

\bibpunct{}{}{,}{s}{}{\textsuperscript{,}}
\begin{document}

\preprint{AIP/123-QED}

\title[Free expansion of a Gaussian wavepacket using operator manipulations]{Free expansion of a Gaussian wavepacket using operator manipulations}
\author{Alessandro M. Orjuela}
\affiliation{Department of Physics, Bowdoin College, Maine USA}
\author{J. K. Freericks}%
\email{james.freericks@georgetown.edu}
\affiliation{ 
Department of Physics, Georgetown University, 37th and O Sts. NW, Washington, DC 20057 USA
}%

\date{\today}

\begin{abstract}
The free expansion of a Gaussian wavepacket is a problem commonly discussed in undergraduate quantum classes by directly solving the time-dependent Schr\"odinger equation as a differential equation. In this work, we provide an alternative way to calculate the free expansion by recognizing that the Gaussian wavepacket can be thought of as the ground state of a harmonic oscillator with its frequency adjusted to give the initial width of the Gaussian, and the time evolution, given by the free-particle Hamiltonian, being the same as the application of a time-dependent squeezing operator to the harmonic oscillator ground state.
Operator manipulations alone (including the Hadamard lemma and the exponential disentangling identity) then allow us to directly solve the problem.
As quantum instruction evolves to include more quantum information science applications, reworking this well known problem using a squeezing formalism will help students develop intuition for how squeezed states are used in quantum sensing.
\end{abstract}

\maketitle
\section{\label{sec:level1}Introduction}

A common example of time evolution taught in quantum mechanics classes---in fact, in many cases, the first example that is considered---is the free expansion of a Gaussian wavepacket starting from a state with variance $\sigma$ and zero average momentum, which is in the following wavepacket at $t=0$:
\begin{equation}
    \psi(x,t{=}0)=\frac{1}{(2\pi\sigma^2)^{\frac{1}{4}}}e^{-x^2/4\sigma^2}.
\end{equation}
It is then solved by transforming from position to momentum space, time evolving in momentum space, and then transforming back to position space. These Fourier transformation integrals are elementary to do, because they are in a Gaussian form and Gaussians integrate to Gaussians. But the analysis is fairly technical and does not provide great physical intuition as to what is happening (aside from the notion that because we have a spread in momentum, this means that components of the wavepacket move in position space at different rates and hence the position-space wavepacket spreads in time). 

In this work, we present a different methodology to solve this problem, based on the modern concept of squeezing. Squeezing is a phenomenon whereby a quantum state maintains a minimum uncertainty product, but the uncertainties in position and momentum, oscillate in time, with each individual uncertainty becoming very small at some moment of time and the other uncertainty becoming very small at a later time. Squeezing is often discussed within the context of a simple harmonic oscillator. Here, the squeezing operator looks similar to the exponentiation of the Hamiltonian itself, but with different coefficients in front of the individual terms. The application to the free-expansion of a Gaussian comes from the facts that the ground state of a harmonic oscillator is also a Gaussian, and that the free time-evolution operator is a squeezing operator (for the harmonic oscillator system). Hence, the time evolution of a Gaussian can be solved by simply applying a squeezing operator to the harmonic oscillator ground state!

Squeezed states can be difficult to understand. One way to physically picture them is via a quantum quench. If a system is in a harmonic oscillator energy eigenstate with an initial frequency $\omega_i$, and then suddenly the frequency changes to $\omega_f$, then we would expect the initially steady state to evolve over time. This subsequent time evolution is one way to think of the behavior of a squeezed state. Unlike a coherent state, which preserves its shape as it evolves in time, the squeezed state changes its shape as it evolves, periodically changing from a squeezed state, with smaller variance in position space, to an expanded state, with a larger variance in position space. The period of the shape oscillations is one half the period of the new oscillator. If, the quench goes all the way to $\omega_f=0$, then we will no longer have periodic motion. Instead, the Gaussian will simply expand forever as a function of time. This is precisely the scenario for the free expansion of a Gaussian that we consider here.

We have surveyed 25~ different undergraduate textbooks to see whether they cover the free-expansion of a Gaussian wavepacket and where this material appears relative to discussions of the operator-based solution of the simple harmonic oscillator. We also looked at whether these textbooks discuss coherent states and squeezed states.  The results are summarized in Table~\ref{tab:books}.   Clearly, none of these texts are ideal for using the material presented here---supplemental material will be required for all of them.

\begin{table*}[t!]
    \centering
    \begin{tabular*}{0.995\textwidth}{l l|c c c c c}
    \hline
    Author& Title& Spreading & SHO in& SHO before & Coherent & Squeezed\\
    & & Gaussian& operator form&Gaussian&states&states\\
    \hline
        Banks & \textit{Quantum Mechanics} &X & X & & X & \\
        Beck  & \textit{Quantum Mechanics} &X  & X & & X & X \\
        Binney and Skinner & \textit{The Physics of Quantum Mechanics} &X  & X & & &  \\
        Brandsden and Joachain& \textit{Quantum Mechanics}& X & & & & \\
        Burkhardt and Leventhal& \textit{Foundations of Quantum Physics}& X & X & & X & \\
        Cohen-Tannoudji, Diu, and Lalo\"e&\textit{Quantum Mechanics} & X & X & & X & \\
        Dicke and Wittke& \textit{Introduction to Quantum Mechanics}&  & X & & & \\
        French and Taylor & \textit{Introduction to Quantum Physics} & X & & & & \\
        Griffiths&\textit{Introduction to Quantum Mechanics}& X & X & X & X & \\
        Hannabuss& \textit{An Introduction to Quantum Theory}&  & X & & X & X \\
        Kroemer& \textit{Quantum Mechanics for Engineering}& X & & & & \\
        Liboff& \textit{Introductory Quantum Mechanics}& X & X & & & \\
        Mahan&\textit{Quantum Mechanics in a Nutshell}& & x & & X &  \\
        McIntyre, Manogue, and Tate& \textit{Quantum Mechanics}&  & X & & & \\
        Miller& \textit{Quantum Mechanics for Scientists} $\cdots$& X & X & & X & \\
        Ohanian& \textit{Principles of Quantum Mechanics}& X & X & & & \\
        Puri & \textit{Nonrelativistic Quantum Mechanics}& X & X & & X & X \\
        Rae and Napolitano& \textit{Quantum Mechanics}&  & X & & & \\
        Robinett& \textit{Quantum Mechanics: Classical Results, $\cdots$}& X & X & & X & \\
        Saxon& \textit{Elementary Quantum Mechanics}& X  & X & & & \\
        Shankar& \textit{Principles of Quantum mechanics}& X & X & & X & \\
        Townshend&\textit{A Modern Approach to Quantum Mechanics}& X & X & & X & \\
        Winter& \textit{Quantum Physics}&  & X & & & \\
        Zettili& \textit{Quantum Mechanics: Concepts and $\cdots$}& X & X & & & \\
        Zweibach& \textit{Mastering Quantum Mechanics: $\cdots$}& X & X & & X & X\\
        \hline
    \end{tabular*}
    \caption{Summary of different topics, as presented in 25 undergraduate quantum mechanics textbooks. The X indicates that the topic is covered in the corresponding textbook. }
    \label{tab:books}
\end{table*}

For undergraduate classes that employ these textbooks, the methodology we discuss 
fits in best with a course organized so that the abstract operator method is used to solve the harmonic oscillator problem \textit{before} a discussion of the spreading Gaussian wavepacket. However, because all of these textbooks have inadequate resources for these topics, extra instruction on squeezed states of the harmonic oscillator is required. Since quantum mechanics classes are likely to change the material they cover to be more aligned with quantum information science, we anticipate many more classes will provide coverage of coherent and squeezed states and be well positioned to discuss the material we develop here in the near future. Indeed, we find from our textbook survey that more recent books are more likely to cover coherent and squeezed states, supporting this trend.

The pedagogical literature has also discussed the free expansion of a Gaussian wavepacket extensively, but we are not aware of any physics education research work
on this squeezed states approach. Our discussion of previous pedagogical work will not be exhaustive here. Much of the previous work is interested in the question of whether and how the wavepacket expands,\cite{bradford,klein} how its shape evolves over time,\cite{mita,andrews,vandijk} and whether a power-series expansion to the time-evolution operator converges.\cite{klein} Some papers discuss alternative methodologies to compute the evolving wavepacket.\cite{lai}~  We are aware of three works that discuss this problem in relationship to squeezed states of the simple harmonic oscillator.\cite{ford,robinett2,kim}~ Unlike this previous work, our work emphasizes the operator approach to squeezed states. 

The ground-state wavefunction of a simple harmonic oscillator, with mass $m$ and frequency $\omega$ is given by
\begin{equation}
    \psi_{gs}(x)=\langle x|0\rangle=\left (\frac{m\omega}{\pi\hbar}\right )^{\frac{1}{4}}e^{-m\omega x^2/2\hbar}.
    \label{eq:gs}
\end{equation}
Here, $|0\rangle$ is the ground state of the simple harmonic oscillator and $|x\rangle$ is the position eigenstate at the location $x$.
The variance of $\sigma^2$ of the initial Gaussian is set by choosing $\omega=\hbar/2m\sigma^2$. Next, we can use a displacement operator to displace the Gaussian in both position space and momentum space. The displacement by $x_{\scriptscriptstyle{0} }$ and $p_{\scriptscriptstyle{0} }$, respectively, is given by
\begin{equation}
    \hat{D}(x_{\scriptscriptstyle{0} },y_{\scriptscriptstyle{0} })=e^{-\frac{i}{\hbar}(x_{\scaleto{0}{3pt}}\hat{p}-p_{\scaleto{0}{3pt}}\hat{x})},
\end{equation}
where $\hat{x}$ and $\hat{p}$ are the operators of position and momentum, which satisfy the canonical commutation relation $[\hat{x},\hat{p}]=i\hbar$. The displaced Gaussian state can then be represented by
\begin{equation}
    |\psi_{gs};x_{\scriptscriptstyle{0} },p_{\scriptscriptstyle{0} }\rangle=\hat{D}(x_{\scriptscriptstyle{0} },p_{\scriptscriptstyle{0} })|0\rangle;
\end{equation}
our choice for writing the wavepacket in this fashion fixes its initial global phase.

The free-particle evolves according to the free Hamiltonian, given by $\hat{\mathcal H}_{\text{free}}=\hat{p}^2/2m$, so the time evolution of the initial Gaussian state becomes
\begin{equation}
    |\psi_{gs}(t);x_{\scriptscriptstyle{0} },p_{\scriptscriptstyle{0} }\rangle=e^{-\frac{i}{\hbar}\hat{\mathcal H}_{\text{free}}t}\hat{D}(x_{\scriptscriptstyle{0} },p_{\scriptscriptstyle{0} })|0\rangle,
\end{equation}
and the spreading Gaussian wavefunction evolves according to time via
\begin{equation}
    \psi_{gs}(x,t;x_{\scriptscriptstyle{0} },p_{\scriptscriptstyle{0} })=\langle x|e^{-\frac{i}{\hbar}\hat{\mathcal H}_{\text{free}}t}\hat{D}(x_{\scriptscriptstyle{0} },p_{\scriptscriptstyle{0} })|0\rangle.
\end{equation}

It turns out that this is in the form of a squeezed and displaced simple harmonic oscillator state, which can be seen most easily when we express the position and momentum operators in terms of the ladder operators of the simple harmonic oscillator.

The raising and lowering operators for a simple harmonic oscillator are given by
\begin{equation}
    \hat{a}=\sqrt{\frac{m\omega}{2\hbar}}\left (\hat{x}+i\frac{\hat{p}}{m\omega}\right )~~\text{and}~~\hat{a}^\dagger=\sqrt{\frac{m\omega}{2\hbar}}\left (\hat{x}-i\frac{\hat{p}}{m\omega}\right ).
\end{equation}
The inverse relations are
\begin{equation}
    \hat{x}=\sqrt{\frac{\hbar}{2m\omega}}\left (\hat{a}+\hat{a}^\dagger\right )~\text{and}~\hat{p}=\sqrt{\frac{\hbar m\omega}{2}}\frac{1}{i}\left(\hat{a}-\hat{a}^\dagger\right ).
\end{equation}
We can re-express the free-particle Hamiltonian in terms of the ladder operators as 
\begin{equation}
    \frac{\hat{p}^2}{2m}=-\frac{\hbar\omega}{4}\left ( (\hat{a}^\dagger)^2-\hat{a}^\dagger\hat{a}-\hat{a}\hat{a}^\dagger+\hat{a}^2\right ).
    \label{eq:hfree}
\end{equation}
The most general  squeezing operator is written as 
\begin{equation}
    \hat{S}(\xi,\eta)=e^{-\frac{\xi}{2}\left (\hat{a}^\dagger\right )^2+i\frac{\eta}{2}\left (\hat{a}^\dagger\hat{a}+\hat{a}\hat{a}^\dagger\right )+\frac{\xi^*}{2}\hat{a}^2},
\end{equation}
with $\xi$ a complex number and $\eta$ a real number.   When converted to position and momentum operators, this is the most general unitary operator that can be constructed from quadratics in position and momentum; namely, it is of the form $\exp\big (ia\hat{x}^2+ib(\hat{x}\hat{p}+\hat{p}\hat{x})+ic\hat{p}^2\big )$, with the numbers $a$, $b$, and $c$ independent of each other and expressible in terms of $\xi$ and $\eta$.  Using Eq.~(\ref{eq:hfree}), we find that the time-evolution operator $e^{-i\hat{p}^2t/2m\hbar}$ is simply the squeezing operator with
\begin{equation}
    \xi=-\frac{i\omega t}{2} ~~\text{and}~~\eta=-\frac{\omega t}{2}.
\end{equation}
Hence, the most general spreading Gaussian is a displaced-squeezed state, with the magnitudes of both $\xi$ and $\eta$ increasing linearly in time. Note that this squeezing operator is a special case, in which $\xi=i\eta$. In this case, we find it is simpler to express the simplified squeezing operator in the form
\begin{equation}
    \hat{S}_\pm(k)=e^{ik\left (\hat{a}^\dagger\pm\hat{a}\right )^2},
\end{equation}
where $k=\omega t/4$.
Here, the operator we work with for the time evolution is $\hat{S}_-\left (\frac{\omega t}{4}\right )$, because it involves the exponential of the square of the momentum operator; we will find we also need to work with the exponential of a constant times the square of the position operator, which corresponds to $\hat{S}_+(k')$ for a suitably chosen $k'$.

\section{\label{sec:formalism}Formalism and details of the derivation}

If students know about the operator-based solution of the simple harmonic oscillator,  coherent states and squeezed states, then it is natural to also discuss the spreading of a Gaussian wavepacket in the context of this simplified squeezing operator and the displacement operator. We describe the details for how this works next.
What we find to be particularly interesting about this approach is that we use ladder operators from the simple harmonic oscillator in the analysis, even though the quantum state evolves with respect to a free-particle Hamiltonian,
and there is no frequency scale in the free-particle Hamiltonian.
The simple harmonic oscillator 
(and thereby the frequency we will use) 
enters only because the initial wavepacket is a Gaussian wavepacket with an initial variance $\sigma^2$. This tells us the frequency of the simple harmonic oscillator for which this Gaussian is the ground-state wavefunction, and allows us to examine the spreading of the wavepacket in this fashion.  The formal developments in this work are similar to, but different from earlier work on how to determine the displaced-squeezed state wavefunction using operator methods\cite{freericks_squeezed} in that here we must work with $\eta\ne 0$, whereas there, $\eta=0$.

We start with a coherent state $|\alpha\rangle$ with initial position $x_{\scriptscriptstyle{0} }=0$ and some arbitrary initial momentum $p_{\scriptscriptstyle{0} }$. Then, using the standard notation for coherent states, we have that $\alpha=ip_{\scriptscriptstyle{0} }\frac{1}{\sqrt{2\hbar m\omega}}$ for this special case.  We use the free-particle Hamiltonian, with $m$ and $\omega$ fixed in our calculations, and we have that the initial variance of the probability distribution is given by $\sigma_{\scriptscriptstyle{0} }^2=\Delta^2_x=\hbar/2m\omega$, since it is a coherent state.

To obtain the time-dependent wavefunction, we take the overlap of a position bra with the time-evolved initial coherent state;
\begin{equation}
\psi(x,t)=\langle x| e^{-\frac{i}{\hbar}\hat{\mathcal H}_{\text{free}}t}|\alpha\rangle=\langle x|e^{-\frac{i t}{2\hbar m}\hat p^2}e^{\frac{ip_{\scaleto{0}{3pt}}}{\hbar}\hat x}|0\rangle\text{.}\label{eq:start}
\end{equation}
 The coherent state is simply a momentum translation operator applied to the harmonic oscillator ground state. We will also find it convenient to later introduce the position state at the origin  $|0_x\rangle$, which satisfies $\hat{x}|0_x\rangle=0$. Then, using the translation operator for position  yields the position eigenstate at $x$ via $|x\rangle=e^{-\frac{i}{\hbar}x\hat{p}}|0_x\rangle$, so that $\hat{x}|x\rangle=x|x\rangle$.
In addition, because the position translation operator has the semigroup property, given by $e^{-\frac{i}{\hbar}x\hat{p}}e^{-\frac{i}{\hbar}x'\hat{p}}=e^{-\frac{i}{\hbar}(x+x')\hat{p}}$, we find that  $|x+x'\rangle=e^{-\frac{i}{\hbar} x'\hat{p}}|x\rangle=e^{-\frac{i}{\hbar} x\hat{p}}|x'\rangle$ both hold. We will also use this identity below.

Our manipulations will employ two other operator identities: (i) the Hadamard lemma,
\begin{equation}
    e^{\hat{A}}\hat{B}e^{-\hat{A}}=\hat{B}+[\hat{A},\hat{B}]+\frac{1}{2!}[\hat{A},[\hat{A},\hat{B}]]+\cdots
\end{equation}
with the $n$th term in the summation equal to \begin{equation}
    \frac{1}{n!}\underbrace{[\hat{A},[\hat{A},\cdots\hat{B}]\cdots ]}_{\substack{n~\text{nested commutators}\\ \text{of $\hat{A}$ with $\hat{B}$}}}
\end{equation}
and (ii) the exponential disentangling identity, which is discussed in the Appendix.

Now, we begin our work to simplify Eq.~(\ref{eq:start}). We use a multiply by 1 to create a Hadamard lemma expression, which, after evaluating the Hadamard lemma effectively moves the $\exp(\tfrac{ip_0}{\hbar}\hat{x})$ term from the right of the time evolution operator to its left.
\begin{align}
\psi(x,t)&=\langle x|\underbrace{e^{\frac{ip_{\scaleto{0}{3pt}}}{\hbar}\hat x}e^{-\frac{ip_{\scaleto{0}{3pt}}}{\hbar}\hat x}}_{=1}e^{-\frac{i t}{2\hbar m}\hat p^2}e^{\frac{ip_{\scaleto{0}{3pt}}}{\hbar}\hat x}|0\rangle\nonumber\\
&=\langle x|e^{\frac{ip_{\scaleto{0}{3pt}}}{\hbar}\hat x}\,\underbrace{e^{-\frac{ip_{\scaleto{0}{3pt}}}{\hbar}\hat x}e^{-\frac{i t}{2\hbar m}\hat p^2}e^{\frac{ip_{\scaleto{0}{3pt}}}{\hbar}\hat x}}_{\text{Hadamard}}|0\rangle\nonumber\\
&=e^{\frac{ixp_{\scaleto{0}{3pt}}}{\hbar}}\langle x|e^{-\frac{it}{2\hbar m}(\hat p +p_{\scaleto{0}{3pt}})^2}|0\rangle\text{.}\label{eq:second}
\end{align}
Here, we used the fact that $\langle x|\hat{x}=\langle x|x$  to replace $\langle x|\exp(\tfrac{ip_0}{\hbar}\hat{x})$ by $\langle x|\exp(\tfrac{ip_0}{\hbar}x)$ and then moved the number out of the matrix element. Evaluating the Hadamard lemma requires some standard operator manipulations. First, we expand the central exponential in a power series, and note that the exponential factors can be moved into the exponent (this operation is called the braiding relation) because the ``internal'' exponential factors appear in pairs and multiply to 1,
\begin{align}
    e^{-\frac{ip_{\scaleto{0}{3pt}}}{\hbar}\hat x}e^{-\frac{i t}{2\hbar m}\hat p^2}e^{\frac{ip_{\scaleto{0}{3pt}}}{\hbar}\hat x}&=\sum_{n=0}^\infty \tfrac{1}{m!}\left (-\frac{i t}{2\hbar m}\right )^me^{-\frac{ip_{\scaleto{0}{3pt}}}{\hbar}\hat x}\hat {p}^{2m}e^{\frac{ip_{\scaleto{0}{3pt}}}{\hbar}\hat x}\nonumber\\
    &=\sum_{n=0}^\infty \tfrac{1}{m!}\left (-\frac{i t}{2\hbar m}\right )^m\left (\underbrace{e^{-\frac{ip_{\scaleto{0}{3pt}}}{\hbar}\hat x}\hat {p}e^{\frac{ip_{\scaleto{0}{3pt}}}{\hbar}\hat x}}_{\text{Hadamard}}\right )^{2m}\nonumber\\
     &=\sum_{n=0}^\infty \tfrac{1}{m!}\left (-\frac{i t}{2\hbar m}\right )^m\left (\hat {p}-\tfrac{ip_{\scaleto{0}{3pt}}}{\hbar}[\hat{x},\hat{p}]\right )^{2m}\nonumber\\
     &=\sum_{n=0}^\infty \tfrac{1}{m!}\left (-\frac{i t}{2\hbar m}\right )^m\left (\hat {p}+p_{\scaleto{0}{3pt}}\right )^{2m}\nonumber\\
     &=e^{-\frac{i t}{2\hbar m}(\hat{p}+p_{\scaleto{0}{3pt}})^2}.
\end{align}
Note that the Hadamard for the exponent truncates after just the first commutator because it is a number, which subsequently commutes with all operators.

Next, we expand the squared exponent in Eq.~(\ref{eq:second}), pull out the constant terms, let the terms with a linear exponent in $\hat{p}$ act on the bra to translate the position state, and then replace the momentum operator with its expression in terms of ladder operators (of the simple harmonic oscillator) in the quadratic exponent to give us
\begin{align}
\psi(x,t)&=e^{\frac{ip_{\scaleto{0}{3pt}}x}{\hbar}-\frac{itp_{\scaleto{0}{3pt}}^2}{2\hbar m}}\langle x-\tfrac{tp_{\scaleto{0}{3pt}}}{m}|e^{\frac{i\omega t}{4}\left (\hat a^\dagger-\hat  a\right )^2}|0\rangle\text{.}\label{eq:third}
\end{align}
 We need to motivate our next step. The exponent of the operator is proportional to the difference of the ladder operators, and we do not know how such an operator acts on the position bra to the left. But, if we could find a way to change the difference of ladder operators into a sum in the exponent, then the operator in the exponent would be proportional to $\hat{x}^2$, which can be readily applied onto the position eigenbra to its left. This would then leave  us with the ground-state wavefunction evaluated at $x-\tfrac{tp_{\scaleto{0}{3pt}}}{m}$.
In order to accomplish the sign change, we first use the exponential disentangling identity acting on the simple harmonic oscillator ground state and apply it to the operator with a quadratic exponent as given in Eq.~(\ref{eq: disentangling-final-state}). We use the case with a difference of the ladder operators and $k=\omega t/4$. We find Eq.~(\ref{eq:third}) becomes
\begin{equation}
\psi(x,t)=\frac{e^{\frac{ip_{\scaleto{0}{3pt}}x}{\hbar}-\frac{itp_{\scaleto{0}{3pt}}^2}{2\hbar m}}}{\sqrt{1+\frac{i\omega t}{2}}}\langle x-\tfrac{tp_{\scaleto{0}{3pt}}}{m}|e^{\frac{i\omega t}{4+2i\omega t} (\hat a^\dagger)^2}|0\rangle
\text{,}
\end{equation}
where two of the exponential factors in the exponential disentangling identity have been replaced by 1 because $\hat{a}|0\rangle=0$. This means that both $\exp(\beta\hat{a}^\dagger\hat{a})|0\rangle=|0\rangle$ and $\exp(\gamma\hat{a}^2)|0\rangle=|0\rangle$, with $\beta$ and $\gamma$ numbers, as can be verified by expanding the exponentials into power series and acting them on the harmonic oscillator ground-state ket. 
Next, we use Eq.~(\ref{eq: disentangling-final-state}) again, but now for the sum of ladder operators and with $k'=\frac{\omega t}{4(1+i\omega t)}$.  In addition, we start from the expression on the right and replace it by the equivalent expression on the left. A direct calculation gives
\begin{equation}
    \frac{ik'}{1-2ik'}=\frac{i\omega t}{4(1+i\omega t)\left (1-\frac{2i\omega t}{4(1+i\omega t)}\right )}=\frac{i\omega t}{4+2i\omega t}.
\end{equation}

So, we find
\begin{align}
    \psi(x,t)&=\frac{e^{\frac{ip_{\scaleto{0}{3pt}}x}{\hbar}-\frac{itp_{\scaleto{0}{3pt}}^2}{2\hbar m}}}{\sqrt{1+\frac{i\omega t}{2}}}\langle x-\tfrac{tp_{\scaleto{0}{3pt}}}{m}|\sqrt{\frac{1+\frac{i\omega t}{2}}{1+i\omega t}}e^{\frac{i\omega t}{4(1+i\omega t)} \left (\hat a^\dagger+\hat{a}\right )^2}|0\rangle\nonumber\\
    &=\frac{e^{\frac{ip_{\scaleto{0}{3pt}}x}{\hbar}-\frac{itp_{\scaleto{0}{3pt}}^2}{2\hbar m}}}{\sqrt{1+i\omega t}}\langle x-\tfrac{tp_{\scaleto{0}{3pt}}}{m}|e^{\frac{im\omega^2 t}{2\hbar(1+i\omega t)} \hat{x}^2}|0\rangle\nonumber\\
    &=\frac{e^{\frac{ip_{\scaleto{0}{3pt}}x}{\hbar}-\frac{itp_{\scaleto{0}{3pt}}^2}{2\hbar m}}}{\sqrt{1+i\omega t}}e^{\frac{im\omega^2 t}{2\hbar(1+i\omega t)} \left (x-\tfrac{tp_{\scaleto{0}{3pt}}}{m}\right )^2}\langle x-\tfrac{tp_{\scaleto{0}{3pt}}}{m}|0\rangle\nonumber\\
    &=\frac{e^{\frac{ip_{\scaleto{0}{3pt}}x}{\hbar}-\frac{itp_{\scaleto{0}{3pt}}^2}{2\hbar m}}}{\sqrt{1+i\omega t}}e^{\frac{im\omega^2 t}{2\hbar(1+i\omega t)} \left (x-\tfrac{tp_{\scaleto{0}{3pt}}}{m}\right )^2}\psi_{gs}\left (x-\tfrac{tp_{\scaleto{0}{3pt}}}{m}\right ).
\end{align}
Substituting in the ground-state wavefunction from Eq.~(\ref{eq:gs}) yields our
final result
\begin{equation}
    \psi(x,t)=\left(\frac{m\omega}{\pi \hbar}\right )^{\frac{1}{4}}\frac{1}{\sqrt{1+i\omega t}}e^{-\frac{ m\omega}{2\hbar(1+i\omega t)}\left (x-\frac{tp_{\scaleto{0}{3pt}}}{m}\right )^2+\frac{ip_{\scaleto{0}{3pt}}x}{\hbar}-\frac{itp_{\scaleto{0}{3pt}}^2}{2\hbar m}}.
\end{equation}
The probability density becomes
\begin{equation}
|\psi(x,t)|^2=\sqrt{\frac{m\omega}{\pi\hbar(1+\omega^2t^2 )}}e^{-\frac{m\omega}{\hbar(1+\omega^2t^2)}\left (x-\frac{tp_{\scaleto{0}{3pt}}}{m}\right)^2}\text{.}
\end{equation}
Now we see the variance is increasing approximately quadratically in time as
\begin{equation}
\sigma^2(t)=\Delta_x^2(t)=\frac{\hbar}{2m\omega}(1+\omega^2t^2)=\sigma^2(0)(1+\omega^2t^2)\text{.}
\end{equation}

This is, of course, the standard result for the spreading Gaussian wavefunction in position space. Using operator methods to determine it is a straightforward exercise that helps build facility in working with operators. It is also interesting to think of this in terms of squeezing and a ``hidden'' simple harmonic oscillator.

The momentum-space derivation is even simpler. We start with
\begin{equation} 
\phi(p,t)=\langle p|e^{-\frac{i}{\hbar}\hat{\mathcal H}_{\text{free}}t} |\alpha\rangle=\langle p| e^{-\frac{it}{2\hbar m}\hat p^2} e^{\frac{ip_{\scaleto{0}{3pt}}}{\hbar}\hat x}|0\rangle\text{.}
\end{equation}
We can immediately apply the momentum operator onto the momentum eigenstate and replace $\hat{p}\to p$. This gives
\begin{equation} \phi(p,t)=e^{-\frac{it}{2\hbar m} p^2}\langle p|  e^{\frac{ip_{\scaleto{0}{3pt}}}{\hbar}\hat x}|0\rangle\text{.}
\end{equation}
The evaluation of the remaining expectation value is easy once we realize that the exponential operator translates the momentum via $|p\rangle=e^{\tfrac{ip}{\hbar}\hat{x}}|0_p\rangle$, with $\hat{p}|0_p\rangle=0$ defining the momentum state at the origin $|0_p\rangle$. So, we find the exponential operator translates the momentum bra to give us 
\begin{equation}
    \phi(p,t)=e^{-\frac{ip^2t}{2m\hbar}}\langle p-p_{\scaleto{0}{3pt}}|0\rangle.
\end{equation}
The term $\langle p-p_{\scaleto{0}{3pt}}|0\rangle$ is just the ground state of the simple harmonic oscillator in momentum space. It can easily be found  by Fourier transformation, or by operator methods.\cite{rushka}~ Substituting in this result gives us
\begin{equation}
    \phi(p,t)=\frac{1}{(\pi\hbar m\omega)^{\frac{1}{4}}}e^{-\frac{(p-p_{\scaleto{0}{3pt}})^2}{2\hbar m\omega}-\frac{ip^2t}{2m\hbar}},
\end{equation}
which is our final result for the wavefunction. The probability distribution becomes
\begin{equation} |\phi(p,t)|^2=\frac{1}{\sqrt{\pi\hbar m\omega}} e^{-\frac{(p-p_{\scaleto{0}{3pt}})^2}{\hbar m\omega}}\text{.}
\end{equation}
The variance of the momentum-space wavefunction does not change with time:
\begin{equation} 
\sigma_p^2(t)=\Delta_p^2=\frac{\hbar m\omega}{2}\text{.}
\end{equation}

\section{Applications of these identities and new exercises for student learning}

One of the goals in modernizing physics instruction is to bring modern experimental results into the classroom. Recent work on atoms trapped in harmonic potentials has been able to (i) create the ground state, the first excited state, and the second excited state and (ii) has used time-of-flight spectroscopy to measure the momentum distribution of the atom when it is in these different Fock states.\cite{regal}~ The formalism we used to describe the free-expansion of the simple harmonic oscillator can be immediately adapted to analysing this experiment, which makes for a context-rich problem for students studying the expansion of wavepackets. We describe how this can be used in instruction next.

These experiments employ time-of-flight spectroscopy, which is an important experimental technique that does not often get discussed in quantum instruction. The only quantum textbook we are aware of that discusses this technique in any detail is Ballentine's book.\cite{ballentine}~ In fact, even though many textbooks say that one can measure the momentum of a quantum particle, virtually none explain just how one does this. It is subtle, because we usually measure position to \textit{infer} momentum; hence, introducing a discussion of time-of-flight provides a great opportunity to properly explain how Heisenberg's uncertainty principle applies to a real experiment on measuring momentum.

In fact, while many textbooks do describe how one can measure position, they are often silent about how to measure momentum. For example, Griffiths states ``You might wonder how it is enforced in the laboratory---why can't you determine (say) both the position and the momentum of a particle? You can certainly measure the position of the particle, but the act of measurement collapses the wave function to a narrow spike, which necessarily carries a broad range of wavelengths (hence momenta) in its Fourier decomposition.''~ In this discussion, Griffiths is intimating that a simultaneous measurement of position and momentum is impossible.  It is certainly true that \textit{subsequent} measurements of position cannot be used to determine the original momentum, but that does not forbid position and momentum to be measured at the same time, as we see below.  Zweibach explicitly states ``given a particle, you cannot simultaneously know both its position and its momentum.''~ But, a time-of-flight measurement is one way to determine momentum and it does so by measuring position. Is this in conflict with the Copenhagen interpretation? Of course not, because the momentum is inferred from the position measurement and the assumption that the wavepacket is traveling as a free particle. One simply needs to perform the analysis properly. Let us explain carefully how time-of-flight works.

In a time-of-flight experiment, one prepares a reasonably localized state (in Ref.~\onlinecite{regal} the state is localized on the order of $0.3-0.5~\mu$m). Then at a specific time, the harmonic oscillator trapping potential is shut off (a quench to $\omega_f=0$, just like we analyzed above). We start a clock at this time and let the atom expand in its wavepacket for some time before measuring its position (in Ref.~\onlinecite{regal}, the time of flight is about $10~\mu $s). At the moment of the measurement, we know the position to the accuracy of the position detector---but, we also know its momentum, which is inferred from the fact that it traveled as a free particle from its localized origin to where it was detected in the time of flight. So, at the moment of the measurement, we know the position and momentum of the quantum particle for this one experimental shot! 

Does this violate uncertainty? Of course not. The uncertainty principle applies to the spread in the results if we repeat the measurement many times (which will actually determine the momentum distribution). The momentum distribution is that of the original trapped atom, plus any spread in the distribution that occurred due to the time of flight. But, because the momentum distribution does not spread in time, we do really measure the momentum distribution of the atom in its trap. And what about the original position uncertainty? Unfortunately, we cannot determine that, because this measurement is not sensitive to the initial position. But the uncertainty in knowing just what the initial position is (due to the initial wavepacket spread) does lead to an intrinsic uncertainty in the momentum measurement in a time-of-flight experiment, but one can make that uncertainty as small as desired by simply increasing the time for the time of flight (which reduces the relative error for the momentum measurement).

Hence, the time-of-flight measurement does not fall into the standard measurement paradigm taught in most quantum textbooks, and it avoids the uncertainty principle restrictions on individual experimental shots due to the way it is set-up. The time of flight is employed to magnify the position values so that we can distinguish these different positions (due to the magnification effect) and then \textit{infer} the different momenta in the initial momentum distribution.

What gets confused about quantum mechanics postulates is that every quantum state has complementary uncertainties that must obey the uncertainty principle. In other words, no quantum state can be created that has definite position and momentum. And similarly, no von Neumann style projective measurement can measure both position and momentum without obeying uncertainty as well. But the von Neumann paradigm is not sufficient to be used to describe all of the parts of the time-of-flight measurement---it only describes the position measurement part. This is a subtle point, but an important one to convey to students.

So, while one can certainly say one cannot create a quantum state with definite position and momentum, one should never say that one cannot \textit{measure} position and momentum at the same time. One can. And discussing how this works provides a great opportunity to carefully explain the subtleties of the Heisenberg uncertainty principle and the strategies used to measure momentum. These are two topics often given ``short shrift'' in the traditional quantum mechanics courses.

Back to the experiment in Ref.~\onlinecite{regal}. In those measurements, they examine not just the momentum distribution of the ground state, but of the two lowest excited states as well. We can easily generalize the techniques we developed above to determine the momentum distributions for these different Fock states too, and this is an excellent exercise for students post instruction. Since the momentum distribution of a harmonic oscillator is identical in shape to the position distribution, this also can be thought of as a magnified visualization of the original position distribution as well. The calculation is quite simple to do, and this is because all harmonic oscillator Fock states maintain their shape (in momentum space) during time of flight.\cite{vandijk}~ So, if I start in the $n$th Fock state, given by $\tfrac{\left (\hat{a}^\dagger\right )^n}{\sqrt{n!}}|0\rangle$, then the momentum distribution after a time of flight of duration $t$ is simply given by
\begin{equation}
    \phi_n(p,t)=\langle p|e^{-\tfrac{i}{\hbar}\hat{\mathcal{H}}_{free}t}\tfrac{\left (\hat{a}^\dagger\right )^n}{\sqrt{n!}}|0\rangle,
\end{equation}
which can be immediately solved to be
\begin{equation}
    \phi_n(p,t)=\frac{1}{(\pi\hbar m\omega)^{\tfrac{1}{4}}}\frac{1}{\sqrt{n!2^n}}H_n\left (\tfrac{p}{\sqrt{\hbar m\omega}}\right )e^{-\tfrac{p^2}{2\hbar m\omega}-\tfrac{ip^2t}{2\hbar m}}.
\end{equation}
Here, we have introduced the physicist's form of the Hermite polynomial. Details for how to do this using operators has been discussed in Rushka and Freericks.\cite{rushka}

One might say that because we directly measure the position in the time-of-flight, we should really calculate the position distribution at time $t$. One can do this as well for the Fock states, but it takes a few more steps. Again, we assume the initial momentum is zero because we are in an energy eigenstate. Then, the only difference in the derivation is that at the point when we use the exponential disentangling identity, the exponentials of $\hat{a}^2$ and of $\hat{a}^\dagger\hat{a}$ cannot immediately act on the ground state because the $\left (\hat{a}^\dagger\right )^n$ term is in the way. But this causes no problem, as one can move the exponentials past that operator by using a ``multiply-by-one'' followed by the Hadamard lemma. We have to do this twice---once when we use the disentangling identity on the square of the momentum operator and once when we bring the required factors needed to convert the exponential to the square of the position operator. This requires a fair amount of algebra, which then yields
\begin{align}
    \psi_n(x,t)&=\frac{1}{\sqrt{1+i\omega t}}e^{\frac{im\omega^2 t}{2\hbar(1+i\omega t)} x^2}\frac{1}{\sqrt{n!}(1+i\omega t)^n}\nonumber\\
    &\times \langle x|\left (\hat{a}^\dagger+(i\omega t)^2\hat{a}\right )^n|0\rangle.
\end{align}
If $n=0$, this is the same answer as we had before. When $n=1$, we obtain
\begin{align}
    \psi_1(x,t)&=\left (\frac{m\omega}{\pi\hbar}\right )^{\tfrac{1}{4}}\frac{1}{\sqrt{1+i\omega t}}\sqrt{\frac{2m\omega}{\hbar}}\frac{x}{(1+i\omega t)}\nonumber\\
    &\times e^{-\frac{m\omega}{2\hbar(1+i\omega t)} x^2}.
\end{align}
The first excited state probability distribution then becomes
\begin{equation}
    |\psi_1(x,t)|^2=\sqrt{\frac{m\omega}{\pi\hbar}}\frac{1}{\sqrt{1+\omega^2t^2}}\frac{2m\omega x^2}{\hbar(1+\omega^2t^2)}e^{-\frac{m\omega}{\hbar(1+\omega^2t^2)}x^2}.
\end{equation}
This has the same shape as the first excited state, but with $x^2\to x^2/(1+\omega^2t^2)$ (and the required change in the normalization factor), as we expect it to.\cite{vandijk}~ The calculation of higher-order excited states follows in a similar fashion, but is tedious to work out the details, because the Hermite polynomial must have each power of $x$ scaled with a different factor, leading to cumbersome equations to work out. So we stop here.

\section{Conclusion}

In this work, we showed an alternative way to determine the expansion of a Gaussian wavepacket in free space as a function of time by mapping the problem onto the application of a squeezing operator on a simple-harmonic-oscillator ground state with the same variance as the initial wavepacket at $t=0$. The squeezing operator used here is particularly simple, which allows us to use an exponential disentangling identity in the derivation that is easy to derive. Given the increasing interest in quantum information science, especially in quantum sensing, we feel this approach will provide an interesting new perspective on an old problem. 

We worked with operators throughout because we believe working with operators makes the material accessible to more students since it does not require calculus (calculus is only needed to determine normalization constants, and those can be told to the student if needed). 

All of the pedagogy used with the free-expansion of a Gaussian wavepacket can also be used if it is taught this way. This pedagogy is well-known and appears in many textbooks, so we do not rehash it here.
Instead, we described how this approach can be employed to introduce a modern  time-of-flight experiment that measures the momentum distribution of a quantum particle in a harmonic trap and in the ground state or a low-lying excited state. This provides an interesting new application of the freely expanding Gaussian, which should become a standard in quantum mechanics courses of the future.

\section{Acknowledgments}
J.K.F. was supported by the National Science Foundation under Grant No. PHY-1915130. A.M.O. was supported by the National Science Foundation under grant number Grant No. DMR-1950502 as an REU student. In addition, J.K.F. was supported by the McDevitt bequest at Georgetown University. 

\section{Author Declarations}
The authors have no conflicts to disclose.

\setcounter{equation}{0}
\renewcommand{\theequation}{A\arabic{equation}}
\section{Appendix: Simplified Exponential Disentangling Identity for the Symplectic Group}
We want to find a disentangled expression for
\begin{equation} 
e^{ik\left (\hat{a}^\dagger\pm\hat{a}\right)^2}=e^{ik\big((\hat a^\dagger)^2\pm(\hat a ^\dagger\hat a+\hat a\hat a^\dagger)+\hat a^2\big)}  \text{,}
\end{equation}
with $k$ a complex number.
Disentangling identities are proven by mapping the operators in the exponent onto Lie algebra generators.\cite{disentangling_identity}~ Here, the Lie algebra is that of the symplectic group. We employ a faithful matrix representation of that Lie algebra to derive an identity of the exponential operator in terms of the product of three different exponential operators---each involving one of the Lie algebra generators. Then, because the representation is faithful, the identity holds when expressed in terms of the Lie generators as well.

We start by working with the following operators, which serve as generators of the symplectic Lie algebra:
\begin{equation} 
\hat K_0=\frac{1}{4}(\hat a^\dagger\hat a+\hat a\hat a^\dagger), ~\hat K_+=\frac{1}{2}(\hat a^\dagger)^2,~\text{and}~ \hat K_-=\frac{1}{2}\hat a^2 \text{.}\end{equation}
They satisfy the commutation relations
\begin{equation}
    \left [\hat{K}_0,\hat{K}_\pm\right ]=\pm\hat{K}_\pm~\text{and}~\left [\hat{K}_+,\hat{K}_-\right ]=-2\hat{K}_0,
    \label{eq: commutator-algebra}
\end{equation}
as can be directly worked out from the fact that $[\hat{a},\hat{a}^\dagger]=1$.
This algebra looks quite similar to the more familiar SU(2) algebra, except for the change in sign on the last commutator.
Then, we have
\begin{equation} 
e^{ik\left (\hat a^\dagger\pm\hat a \right)^2}= e^{2ik\left (\hat K_+\pm 2\hat K_0+\hat K_-\right )}\text{.}
\end{equation}
and our goal is to find $a$, $b$, and $c$ such that
\begin{equation} e^{2ik(\hat{K}_+\pm 2\hat{K}_0+\hat{K}_-)}=e^{a\hat{K}_+}e^{b\hat{K}_0}e^{c\hat{K}_-}\text{.}
\label{eq: disentangling1}
\end{equation}
We create a faithful representation of the Lie algebra via the following $2\times 2$ matrix representation of the Lie algebra:
\begin{equation} \hat{K}_0\leftrightarrow\frac{1}{2}\begin{pmatrix*}[r] -1&0\\0&1\end{pmatrix*},~\hat{K}_+\leftrightarrow\begin{pmatrix*}[r]0&0\\-1&0\end{pmatrix*},~\text{and}~\hat{K}_-\leftrightarrow\begin{pmatrix*}[r]0&1\\0&0\end{pmatrix*}\text{.}
\end{equation}
It is a worthwhile exercise to show that these matrices satisfy the symplectic Lie group commutator algebra in Eq.~(\ref{eq: commutator-algebra}). Using this representation, we find that
\begin{equation}
    \hat{K}_+\pm 2\hat{K}_0+\hat{K}_-\leftrightarrow \begin{pmatrix*}[r]\mp1&1\\-1&\pm1\end{pmatrix*}=M_\pm.
\end{equation}
The matrices $M_\pm$ are nilpotent with index two---in other words, $M_\pm^2=0$. Then the power-series expansion of the exponential truncates after the first term, so we have
\begin{equation}
    e^{2ikM_\pm}=\mathbb{I}_2+2ikM_\pm=\begin{pmatrix}1\mp 2ik&2ik\\-2ik&1\pm 2ik\end{pmatrix}.
\end{equation}
Each term in the right-hand side of Eq.~(\ref{eq: disentangling1}) can also be easily calculated, and we find
\begin{align}
    e^{a\hat{K}_+}e^{b\hat{K}_0}e^{c\hat{K}_-}&\leftrightarrow \begin{pmatrix*}[r]1&0\\-a&1\end{pmatrix*}\begin{pmatrix*}[c]e^{-\frac{b}{2}}&0\\0&e^{\frac{b}{2}}\end{pmatrix*}\begin{pmatrix}1&c\\0&1\end{pmatrix}\nonumber\\
    &=\begin{pmatrix}e^{-\frac{b}{2}}&ce^{-\frac{b}{2}}\\-ae^{-\frac{b}{2}}&-ace^{-\frac{b}{2}}+e^{\frac{b}{2}}\end{pmatrix}.
\end{align}
We equate these two matrices to find that
\begin{equation}
    a_\pm=c_\pm=\frac{2ik}{1\mp2ik}~\text{and}~b_\pm=-2\ln(1\mp2ik).
\end{equation}
Hence, the exponential disentangling identity becomes
\begin{equation}
    e^{ik\left (\hat{a}^\dagger\pm\hat{a}\right)^2}=e^{\frac{ik}{1\mp 2ik}\left (\hat{a}^\dagger\right)^2}e^{-\frac{1}{2}\ln(1\mp2ik)\left (\hat{a}^\dagger\hat{a}+\hat{a}\hat{a}^\dagger\right)}e^{\frac{ik}{1\mp 2ik}\hat{a}^2}.
    \label{eq: disentangling-final}
\end{equation}
Acting this on the simple-harmonic-oscillator ground state then gives us
\begin{equation}
    e^{ik\left (\hat{a}^\dagger\pm\hat{a}\right)^2}|0\rangle=\frac{1}{\sqrt{1\mp 2ik}}e^{\frac{ik}{1\mp 2ik}\left (\hat{a}^\dagger\right)^2}|0\rangle
    \label{eq: disentangling-final-state},
\end{equation}
because $\hat{a}|0\rangle=0$.
This last identity is used in our derivation with two different $k$ values, $k=\tfrac{\omega t}{4}$ and $k'=\tfrac{\omega t}{4(1+i\omega t)}$.
\\
\rule{0.48\textwidth}{1.2pt}

\pagebreak
\end{document}